\documentclass[submission,copyright,creativecommons]{eptcs}
\providecommand{\event}{Post Proceedings of ADG'23} 
\usepackage[T1]{fontenc}
\usepackage{xurl,amsmath}
\usepackage{cite}
\usepackage{graphicx,subfigure,url}
\usepackage{amsfonts,txfonts}
\usepackage{color}
\usepackage{booktabs}
\usepackage{amssymb}
\usepackage{doi}
\usepackage{hyperref}
\usepackage{iftex}

\title{3D Space Trajectories and beyond:\\ Abstract Art Creation with 3D Printing}
\author{Thierry Dana-Picard
\institute{Jerusalem College of Technology\\ Jerusalem, Israel}
\email{ndp@jct.ac.il}
\and
Matias Tejera \qquad\qquad Eva Ulbrich
\institute{Johannes Kepler University\\
Linz, Austria}
\email{\quad Mathias.Tejera@jku.at \quad\qquad Eva.Ulbrich@jku.at}
}
\def\titlerunning{3D Space Trajectories, Art and 3D Printing}
\def\authorrunning{T. Dana-Picard, M. Tejera \& E. Ulbrich}

\begin{document}
\maketitle

\begin{abstract}
We present simple models of trajectories in space, both in 2D and in 3D. The first examples, which model bicircular moves in the same direction, are classical curves (epicycloids, etc.). Then, we explore bicircular moves in reverse direction and tricircular moves in 2D and 3D, to explore complex visualisations of extraplanetary movements. These moves are studied in  a plane setting. Then, adding increasing complexity, we explore them in a non planar setting (which is a closer model of the real situation). The exploration is followed by using these approaches for creating mathematical art in 2D and 3D printed objects, providing new ways of mathematical representations. Students’ activities are organized around this exploration.
\end{abstract}

\section{Introduction}
All over the world, newspapers and TV news are full of reports about launching satellites, the International Space Station, the Chinese space station, Mars exploration and the Artemis project to establish a permanent human presence on the Moon. Nowadays,
the NASA offers the public to send their names on a probe to be launched in 2024 and arrive to Encelade, an icy moon of Jupiter, in 2030. With such an ubiquitous topic, students asked a lot of questions, about spacecrafts, their trajectories, their trajectories,  why these are curved and sometimes complicated, etc. Numerous dedicated websites are freely accessible, showing representations of trajectories of extraplanetary objects. These are connected to the students' cultural background, on which it is worth to rely in order to attract students to mathematics, and to show applications in real world \cite{cultural-background}. This paper explores mathematical situations with a STEAM approach visualising curves in 2D and 3D with various technologies to use the motivational fascination of outer space from students to connect to mathematical modelling.

When asking about spacecrafts, they wish to understand the trajectories. Not all the news items include graphs and maps of the trajectories, but they frequently do so and can be the source of questions, whence of mathematical activities. These are good reasons for mathematics educators to be part of this atmosphere, showing complex real world applications of mathematics.
Examples could be curves describing trajectories or calculating the speed of objects in an accessible way by interactive visualizations and explanatory
animations. Students have the opportunity to create and explore these trajectories themselves by visualising them using mathematical modelling and certain technologies and we present possible approaches in 2D and 3D.

\begin{figure}[htb]
\begin{center}
\scalebox{0.4}{
\includegraphics{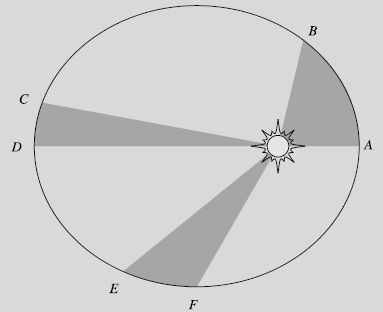}
}
\caption{Kepler's $2^{\text{nd}}$ law of planetary motion}
\label{fig Kepler 2nd law}
\end{center}
\end{figure}

According to the $1^{\text{st}}$ Kepler's law (see \cite{astronomy}, p. 127), the orbit of a planet around the Sun is an ellipse, with the Sun at one of the foci. As the foci are very close, actually both inside the Sun,\footnote{Actually, in a system of two objects, both orbit their common center of gravity. The system Sun-Earth's center of gravity is inside the Sun, therefore considering the Earth as orbiting the Sun is acceptable. Of course, every other pair Sun-Planet presents the same situation.} in order to make the example as simple as possible, we consider an approximation of the orbits as coplanar concentric circles. Kepler's $2^{\text{nd}}$ law is illustrated by figure \ref{fig Kepler 2nd law}, taken from \cite{astronomy} p. 129: the areas of the shaded sectors, covered by the radius in equal times (i.e. it takes equal times to travel distances $AB$, $CD$ and $EF$), are equal.   

In our simplified model, we consider motion with constant angular velocity on circular orbits. We compute the velocities according to the year length of the planet, with Earth year equal to 1. Note that Table \ref{table orbital data} displays just the eight official planets acknowledged by the international astronomical organization. According to the $3^{\text{rd}}$ Kepler's law, the orbital velocity is a function of the distance to the Sun.

\begin{table}[ht]
\begin{center}
\begin{tabular}{c c c}
\hline
\textbf{Planet} & \textbf{Distance to the Sun (km)} & \textbf{Period (1=terrestrial year)} \\
\hline
Mercury &$57.91 \; 10^6$ km & 0.2408 \\
Venus & $108.2 \; 10^6$ km& 0.6152\\
Earth & $149.6 \; 10^6$ km& 1\\
Mars &$227.9 \; 10^6$ km &1.8808 \\
Jupiter & $778.5 \; 10^6$ km& 11.862\\
Saturn & $1.434 \; 10^9$ km& 29.457\\
Uranus & $2.871 \; 10^9$ km& 84.018\\
Neptune &$4.495 \; 10^9$ km & 164.78\\
\hline
\end{tabular}
\end{center}
\label{table orbital data}
\caption{Some orbital data}
\end{table}
Because of the huge differences between the distances and the hardware constraints\footnote{A general study of constraints, either of the hardware or of the software can be found in \cite{trouche}, with some extension in \cite{dp-constraints}.} (we mean mostly the size of the screen and the number of available pixels), we will consider examples with Earth and Mars only. The same activities can be done with the pair Venus-Earth, they will produce the same family of curves. Note that in order to make the first examples easy, we use approximations less precise than in Table \ref{table orbital data}.

The visualisations we explore in this paper are created by two softwares called GeoGebra and Maple to utilise their respective strengths. We use GeoGebra,\footnote{ Freely downloadable from \url{http://www.geogebra.org}.} whose main characteristic is devoted to Dynamic Geometry. For some applications, including automated determination of loci and envelopes, it can be supplemented with the package GeoGebra-Discovery.\footnote{Look for the last version, freely downloadable from \url{https://github.com/kovzol/geogebra-discovery}.} A general analysis of the
automated methods for loci and envelopes is given in [6]. We will also use the Computer Algebra System Maple for its specific animated affordances, which are different from those of GeoGebra.

\begin{figure}[htb]
\begin{center}
\scalebox{0.4}{
\includegraphics{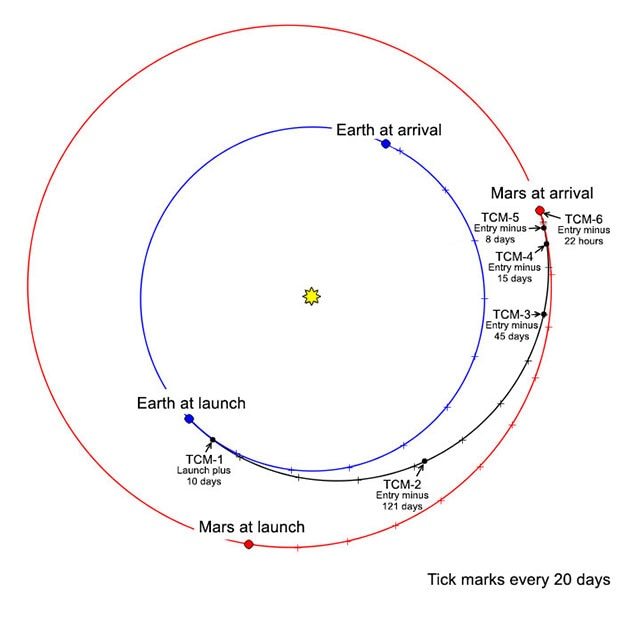}
}
\caption{Trajectory from the Earth to the Mars orbit}
\label{fig Earth to Mars trajectory}
\end{center}
\end{figure}

Exploration of curves obtained as trajectories of points modeling moves in space, such as midpoint or center of gravity of two planets, is described and analyzed in  \cite{virtual,art-in-space}. The present paper is a new contribution, with more complex constructions. Its goal is to present mathematical situations with a STEAM\footnote{STEAM = acronym for Science, Technology, Engineering, Arts and Mathematics.} approach, where plane curves, either algebraic or not, are presented and some of their properties explored using technology. We see this kind of study as an opportunity to connect to the classical families of plane curves in a motivating manner for students and can be used as a unifying frame for cases previously seen as separate cases. Later, space curves given by similar parametric equations are explored, also modelling spatial phenomenon. The ratio of the mean radii of two neighboring planets (such as Venus-Earth, or Earth-Mars) is huge, and still more the ratio
between the mean radius of the Earth orbit around the Sun (about 149 Mkm) and the radius of the Moon’s orbit around the Earth (about 360,000 km), it is
impossible to represent both on a computer screen. Therefore, we chose to work with arbitrary\footnote{By arbitrary, we mean coefficients enabling a representation on the screen, not taken from the orbital data.} coefficients, whose variations provide different curves. We
explore the composition of two circular movements in the same direction (the general case in the Solar System). Figure \ref{fig Earth to Mars trajectory} shows a simplified model of a spacecraft flying to Mars, without explicit presentation of the orbiting direction around Mars.%
\footnote{Credit: NASA/JPL,  \url{https://marspedia.org/File:InSight_Trajectory.jpg}} 

\begin{figure}[htb]
\begin{center}
\scalebox{0.4}{
\includegraphics{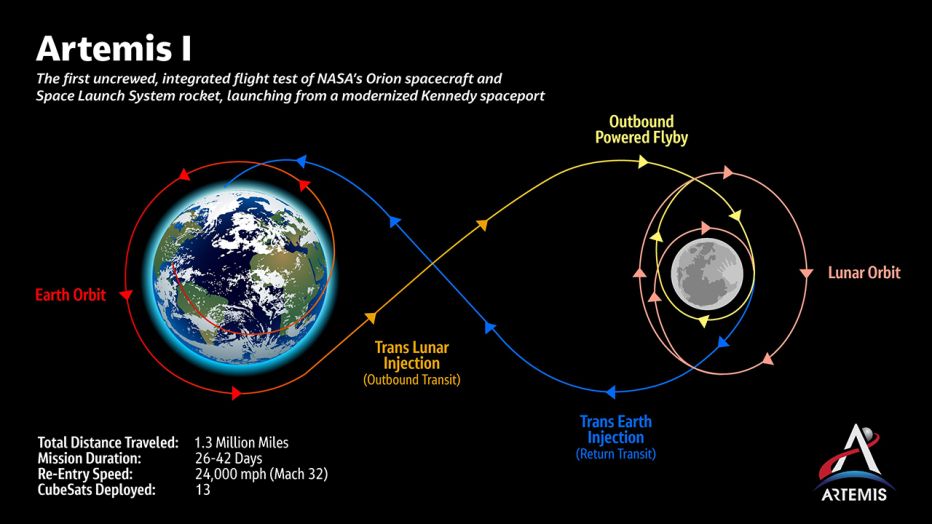}
}
\caption{Artemis orbit to the Moon and around (Credit: NASA)}
\label{fig Artemis trajectory}
\end{center}
\end{figure}

We  also consider the composition of 3 circular movements, inspired by lunar orbiters and observed also from the Sun: they orbit the Moon, which orbits the Earth, which in its turn  orbits the Sun. In Subsection \ref{subsection tricicular movements 1}, we explore the composition of 3 circular movements, all in the same direction. In Subsection \ref{subsection tricicular movements 2}, we explore also models of a composition of movements, two in one direction and the 3rd in reversed direction.  The motivation for this is provided by the trajectories of spacecrafts to the Moon; figure \ref{fig Artemis trajectory} shows a diagram of the trajectory of the Artemis 1 spacecraft, elliptic around the Earth, followed by a transfer orbit made of arcs of ellipses, then elliptic around the Moon in reversed direction.\footnote{The trajectories of future Artemis missions will be different from this one, but based on the same principle.} The geometric locus of the moving object around the Moon, when observed from the Sun, may be an epicycloid, a hypocycloid or another already known curve. Here, more ``exotic'' curves are also explored and plotted; in particular, rotational symmetries of order 7, 11, 13, etc. may be discovered. This provides an opportunity for an interactive exploration of such symmetries.

Jablonski \cite{jablonski} says that ``Mathematical modelling is characterized through its interplay of reality and mathematics.
It offers a way to integrate references to reality into the classroom and shows students where in everyday life their mathematical knowledge can be applied.'' Therefore, we started from real world situations utilizing the amazement created by media reports. The first
examples provide some understanding of how the orbit of the Moon around the Sun looks like, but quickly we explored compositions of movements
without a connection to reality. Changing the parameters (either ratio of radii or ratio of angular velocities) induces important changes on the shape and
topology of the curves. Tricircular moves are inspired by, for example, lunar orbiters (which orbit the Moon, which orbits the Earth, which in its turn orbits the Sun), or Mars orbiters. As already mentioned, figure \ref{fig Earth to Mars trajectory} shows the trajectory of Mars Orbiter, from start to arrival: at first ellipses around the Earth, then a transfer orbit (made of arcs of ellipses), then elliptic orbits around Mars. This can be explained to students. 

Instead of returning from models to the real world situation, which had to be understood, numerous new directions are possible. As an example, curves of degree 8, obtained from a construction disconnected from the physical data, have been explored recently; see \cite{aims}.

Finally, we explore artistic creation using these mathematical models. We obtain curves presenting non usual symmetries and
explore them using our software.

In the real world of the software may change. The exploration of the curves is an important incentive to 3D print them. We quote once again Jablonski \cite{jablonski}: ``The idea of involving real objects in mathematical modelling leads to the question of how much the way in which a real object is introduced might influence the modelling processes of students. Despite its actual physical presence in reality, a real object could be introduced through different representations and provided artefacts, e.g., newspaper articles, photographs, videos, 3D print replications or combinations. Potentially, the different representations of the real object might lead to differences in the modelling activities of students and motivate a comparison of them.''

\section{Classical Curves and beyond}
\label{section clasical curves and beyond}
In all the examples we consider a planet (let us call it the Earth, orbiting the Sun at distance 1 (a reference to 1 astronomical unit, 1 AU) at constant velocity, and completing 1 orbit in 1 year. The other coefficients describe the mean radius of another planet and the length of its own year. The is described by the following parametric presentation:
\begin{equation}
\label{eq Earth}
(x,y)=\left( \cos u, \; \sin u \right),\; u \in \mathbb{R}.
\end{equation}
For the animations with software,  $u \in [-0,2 \pi ]$ is enough with repetitive animation. We denote the parameter by $u$, as in GeoGebra $t$ has a special role.
The second planet is described by
\begin{equation}
\label{eq 2nd planet}
(x,y)=r\; \left (\cos \frac {u}{h}, \; \sin \frac {u}{h} \right),\; u \in \mathbb{R},
\end{equation}
where $r>0$ denotes the radius of the planet's orbit and $h$ encodes the length of its year.

\subsection{Epicycloids in 2D and Extension towards 3D}
\begin{figure}[ht]
\centering
\subfigure[GeoGebra]{\includegraphics[width=8cm]{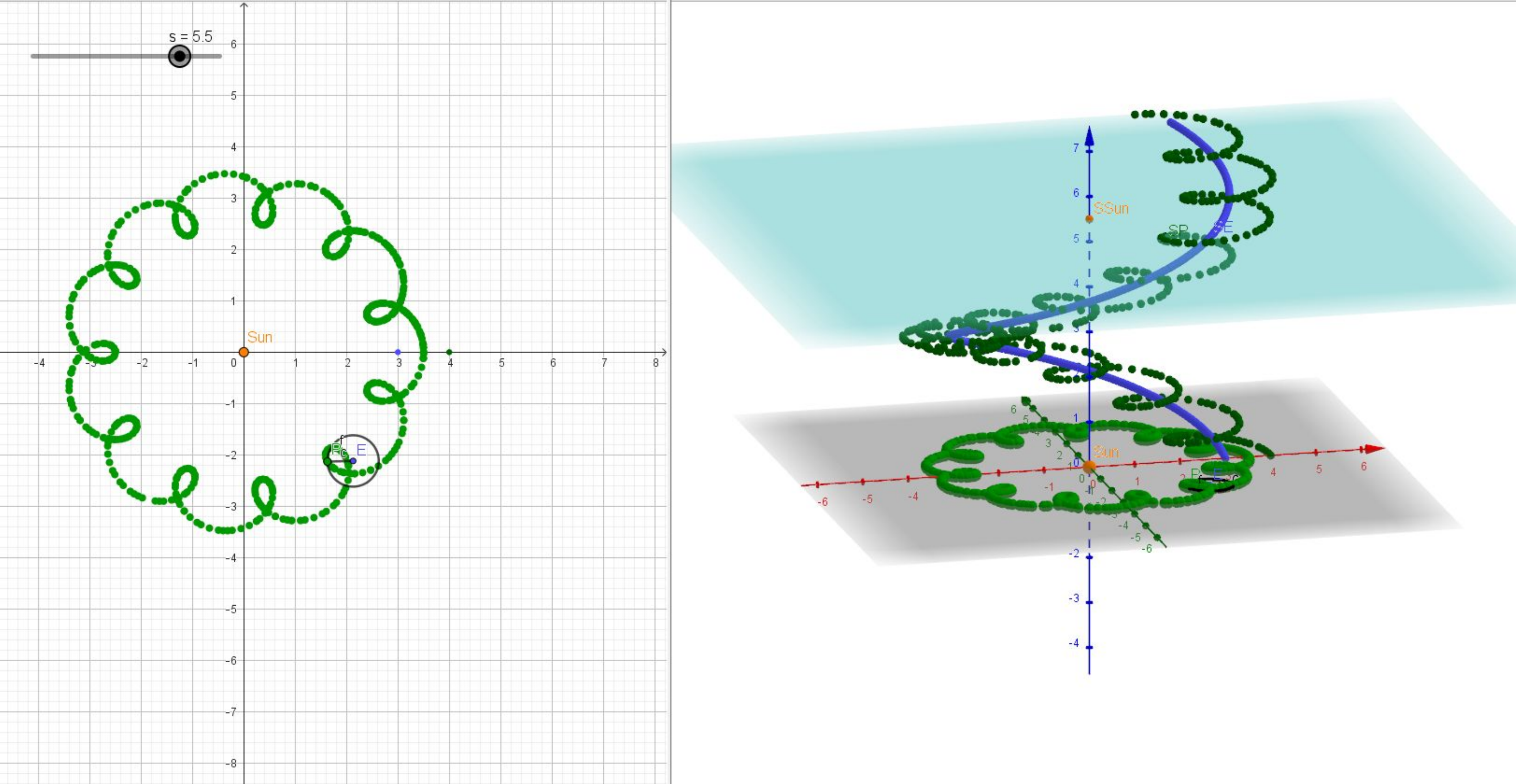}}
\qquad
\subfigure[Maple]{\includegraphics[width=6cm]{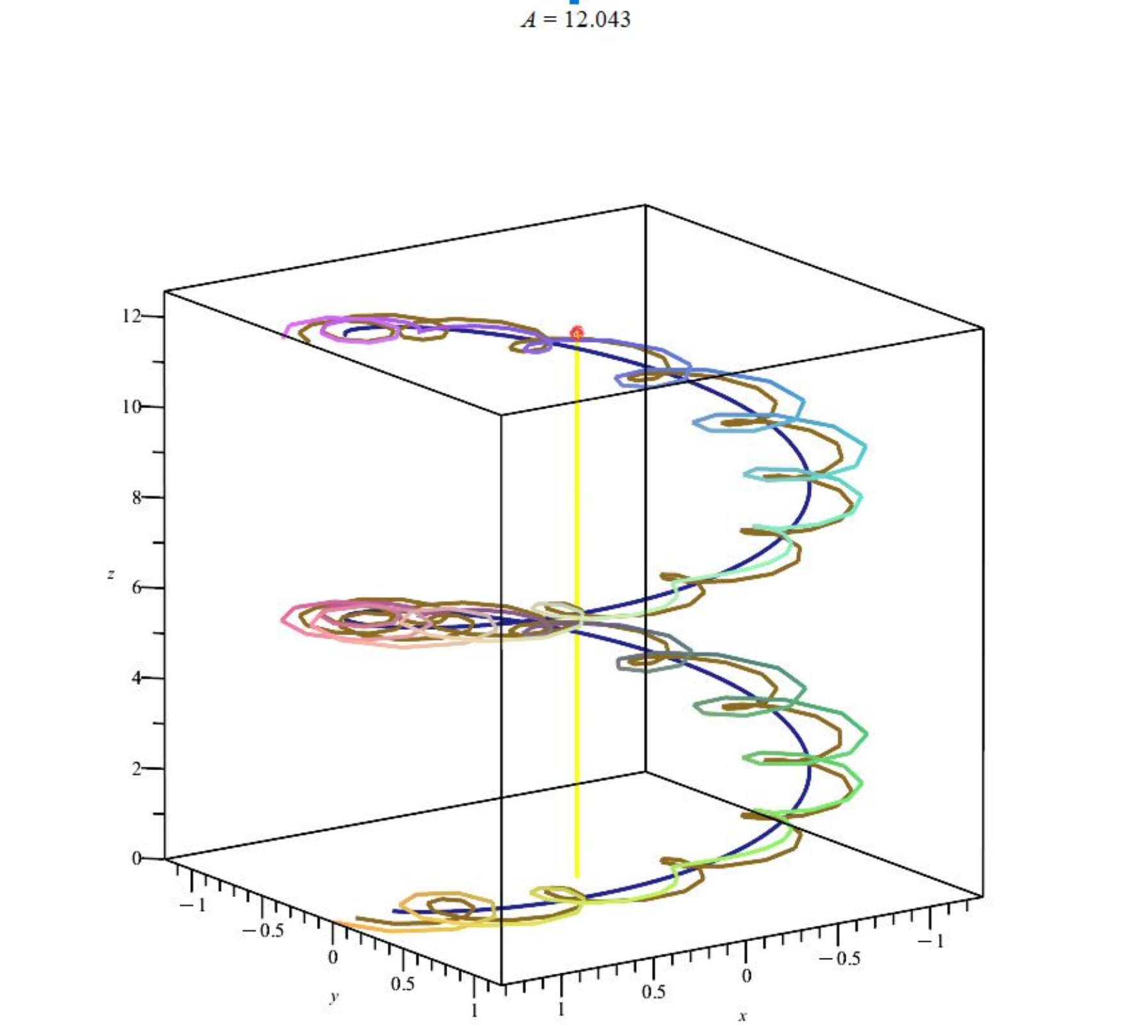}}
\caption{A satellite around a planet orbiting the Sun}
\label{fig sun-planet-moon}
\end{figure}

Figure \ref{fig sun-planet-moon} shows a screenshot of a dedicated GeoGebra applet.\footnote{See \url{https://www.geogebra.org/m/ksyd6hat}.} The parameters can be changed with slider bars. The figure on the left shows the trajectories in the plane containing the Sun, the planet and its satellite. Here the satellite orbits the planet 12 times a year, almost modelling the Moon around the Earth. The figure on the right shows a simulation when the Sun travels on a straight line; note that the 3 objects remain all the time in a plane which moves according to the  Sun.
For the 2D representation, the orbits can be either plotted in a non-animated way using GeoGebra's \textbf{Locus} command or to be animated using the corresponding option of the slider bar. Nite that other way to animate the constructions are available.

As already mentioned, the Sun is also mobile, it has its own orbit. Figure  \ref{fig sun-planet-moon}(a) shows a model where the Sun moves along a segment of line. The planet and its satellite move in a plane containing the 3 objects (this plane is visible in  blue).
The commands are similar in 3D as in 2D. Note that the projection on the plane is on display in the adjacent window. This is due to the total synchronisation of the 3D and 2D windows in GeoGebra.
This task was an incitement to go to 3D printing.

A similar animation can be programmed with Maple. The code is easy: each object is defined in a separate \textbf{plot[animate]} command, them all together they are displayed using the display command. The Sun has two commands: one for plotting a large dot, the other one to plot the trajectory. An animated gif can be obtained with a right-click on the output of the \textbf{display} command. A screenshot is shown in figure \ref{fig sun-planet-moon}(b).
\small
\begin{verbatim}
c1 := spacecurve([cos(t) + 1/5*cos(12*t), sin(t) + 1/5*sin(12*t), t],
      t = 0 .. 4*Pi, thickness = 3, labels = [x, y, z]):
sun := plots[animate](spacecurve, [[0, 0, t], t = 0 .. A], A = 0 .. 4*Pi,
      thickness = 3, color = yellow)
sunplo := plots[animate](pointplot3d, [[0, 0, A]], A = 0 .. 4*Pi,
      color = orange, symbol = sphere)
planet := plots[animate](spacecurve, [[cos(t), sin(t), t], t = 0 .. A],
A = 0 .. 4*Pi, thickness = 3, color = navy)
sat := plots[animate](spacecurve, [[cos(t) + 1/5*cos(12*t), sin(t)
       + 1/5*sin(12*t), t], t = 0 .. A], A = 0 .. 4*Pi,
        color = sienna, labels = [x, y, z]):
display(sun, planet, sat, sunplo)
\end{verbatim}
\normalsize

\subsection{Three Circular Movements with Constant Angular Velocity -- Same Direction}
\label{subsection tricicular movements 1}
\begin{figure}[ht]
\centering
\mbox{
\subfigure[]{\includegraphics[width=4.5cm]{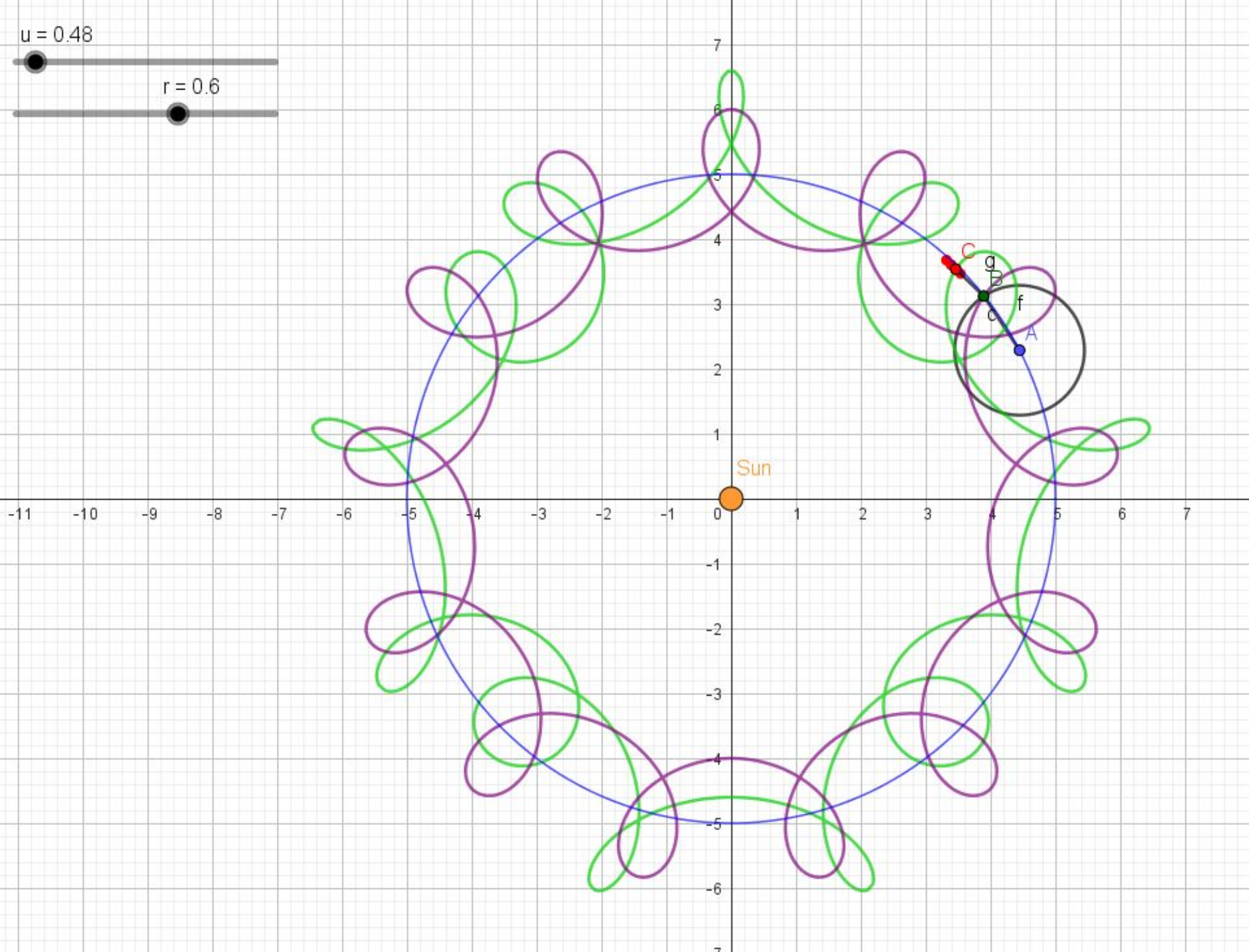}}
\quad
\subfigure[]{\includegraphics[width=4.5cm]{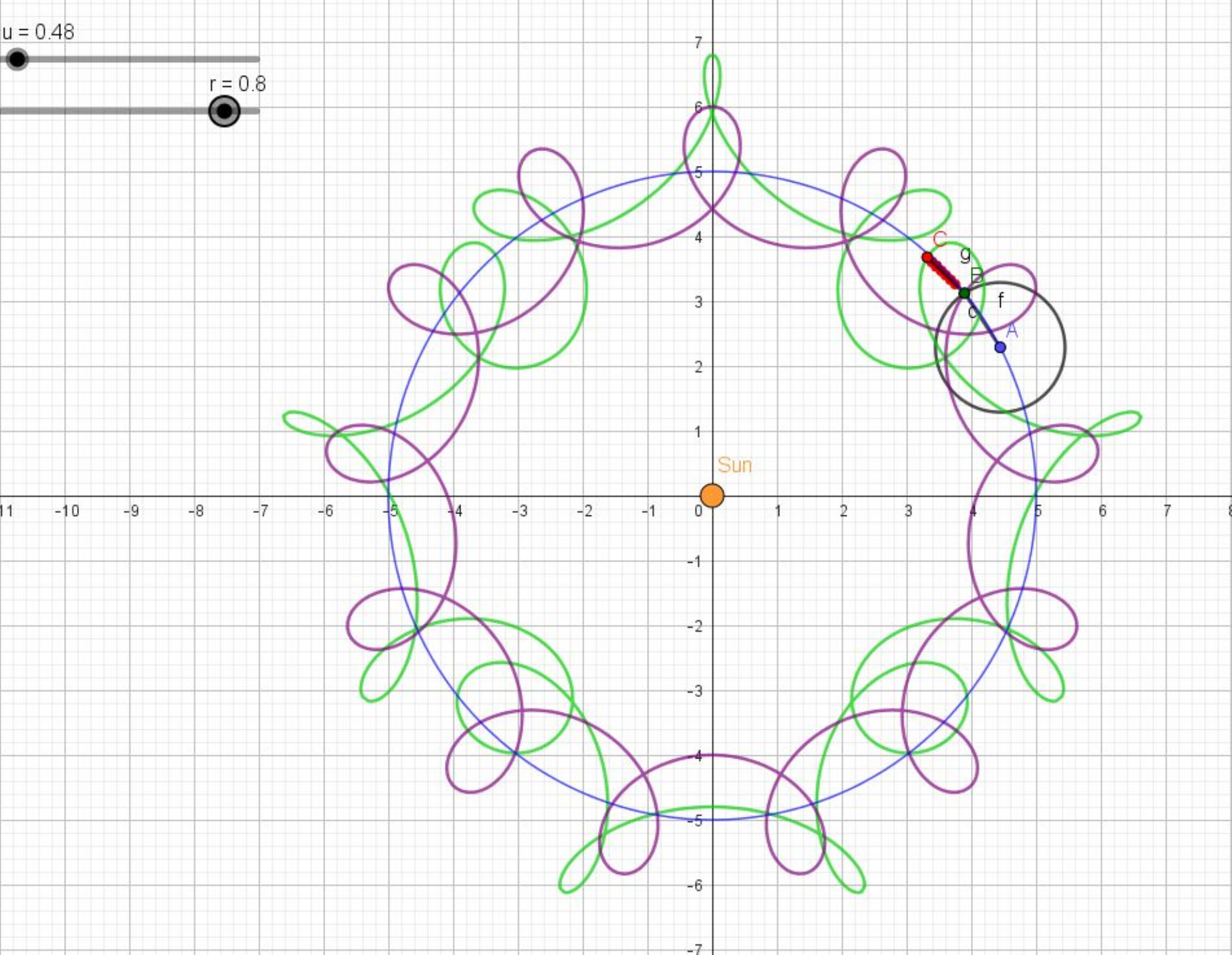}}
\quad
\subfigure[]{\includegraphics[width=4.5cm]{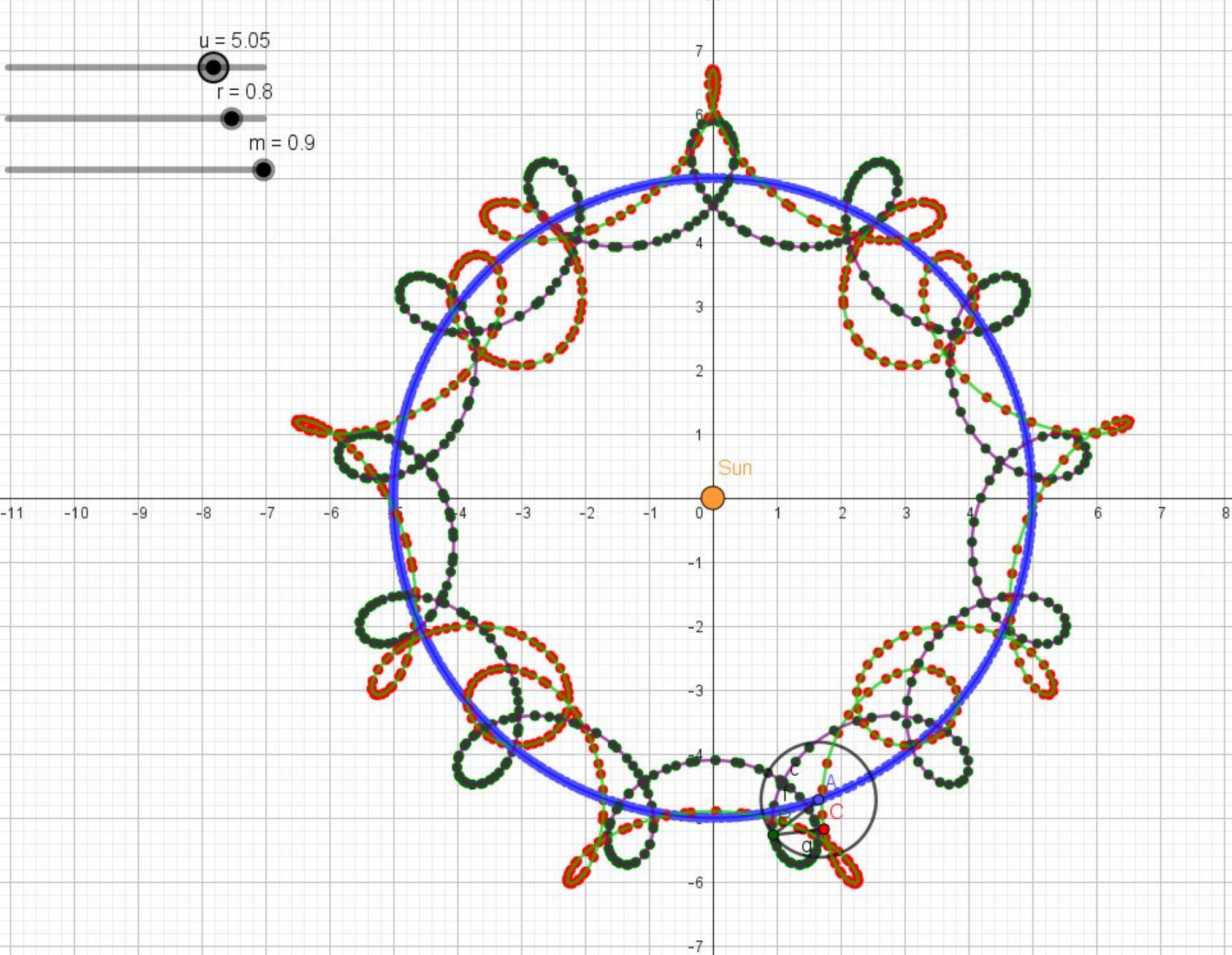}}
}
\caption{Screenshots of a tricircular motion in the same direction}
\label{fig tricircular motion - same direction}
\end{figure}

Figure \ref{fig tricircular motion - same direction} shows snapshots of GeoGebra sessions based on the \textbf{Locus} command. Subfigure (c) is a snapshot of a GeoGebra applet\footnote{See \url{https://www.geogebra.org/m/sagpjzzb}.} with 2 parameters encoding the distances. A further step consists in adding parameters to change the ratios of angular velocities.

\subsection{Three Circular Movements with Constant Angular Velocity -- One in Reverse Direction}
\label{subsection tricicular movements 2}
\begin{figure}[ht]
\centering
\mbox{
\subfigure[A 4-star]{\includegraphics[width=4.5cm]{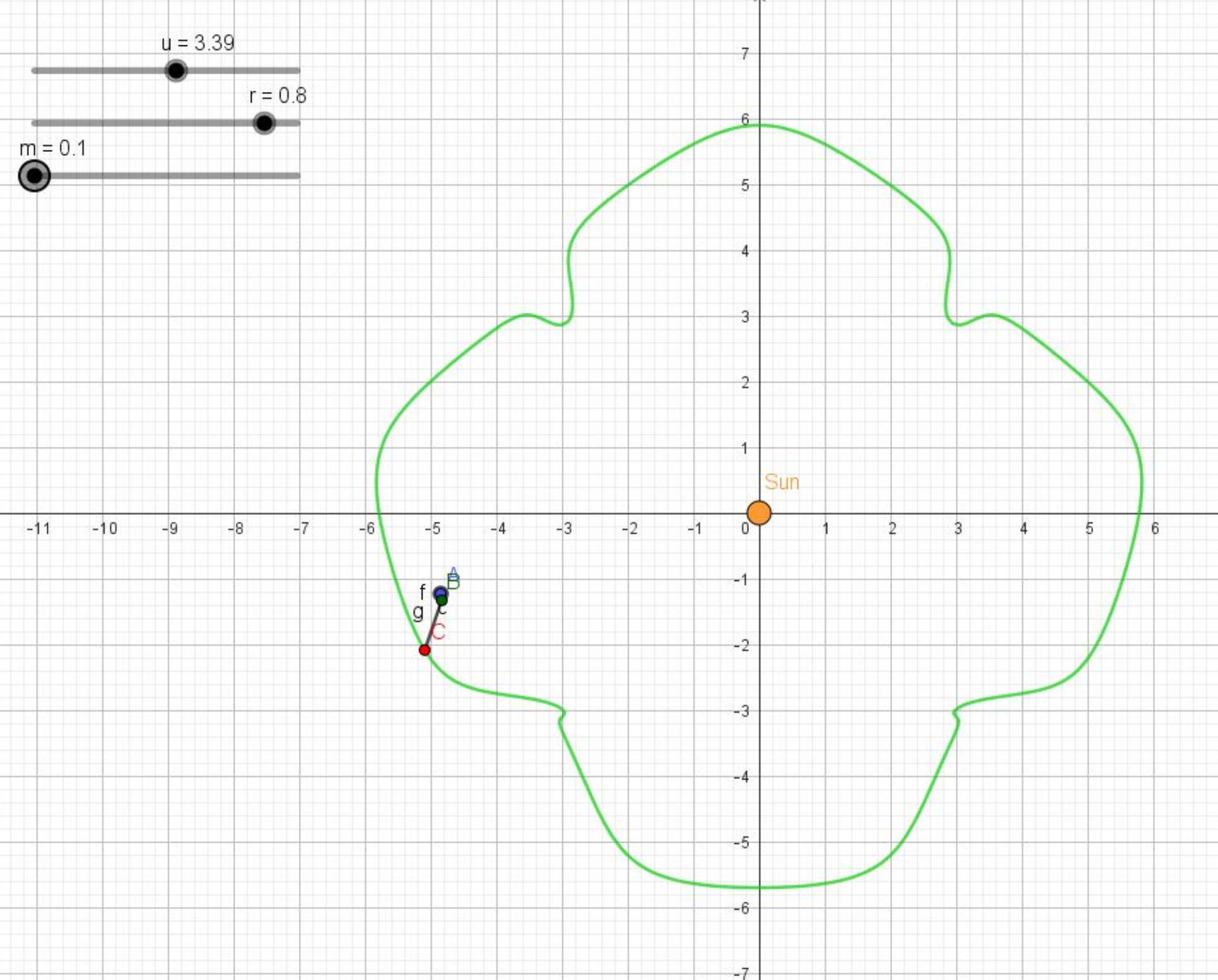}}
\quad
\subfigure[A strange star]{\includegraphics[width=4.5cm]{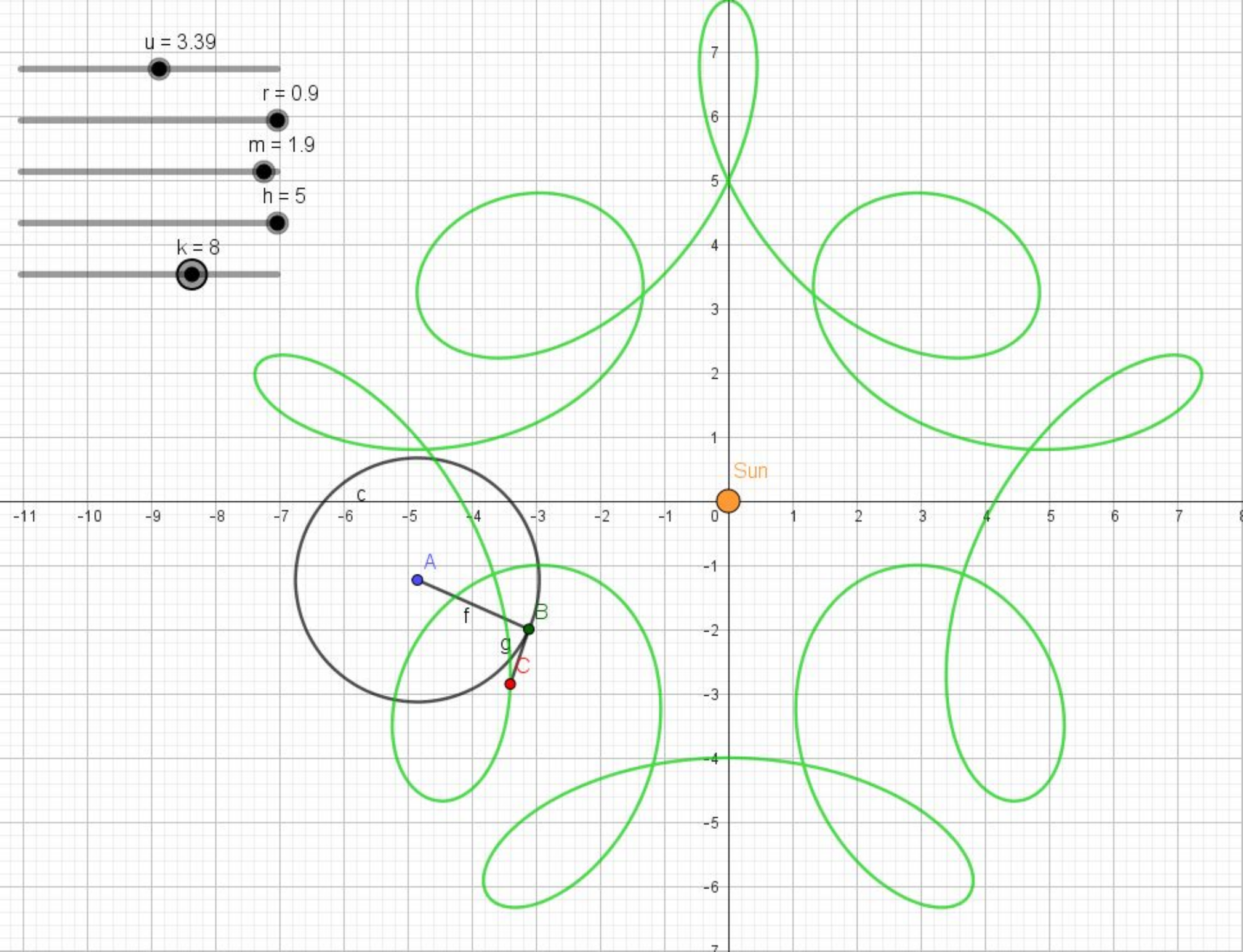}}
\quad
\subfigure[A bat curve]{\includegraphics[width=4.5cm]{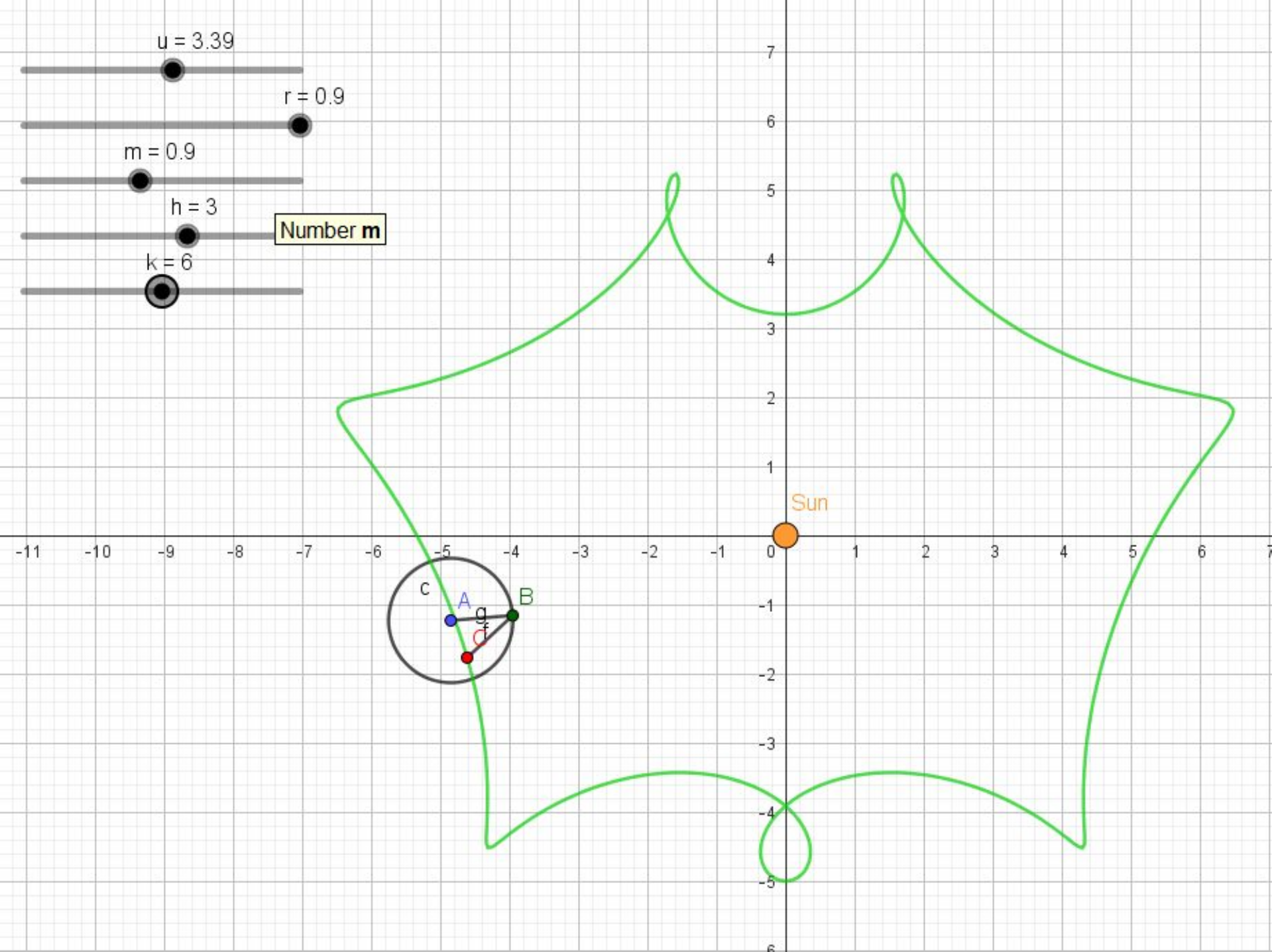}}
}
\caption{Screenshots of a tricircular motion with the middle in reversed direction}
\label{fig tricircular motion - middle direction reversed}
\end{figure}

Figure \ref{fig tricircular motion - middle direction reversed} displays 3 curves obtained with the \textbf{Locus} command, in an applet\footnote{See \url{https://www.geogebra.org/m/xgrx7ntx}.} where all the parameters can vary. In what follows, we explore the symmetries of the obtained curves. These symmetries are often of odd order, a situation which is not frequent in classroom.

We consider now the family of  curves whose parametric equations are as follows:
\begin{equation}
\label{eq multicolor curves}
(x,y,z)=(\cos u,\sin u) + \frac 13 (\cos au, \sin au) +\frac 12 (\sin bu, c\cos bu),
\end{equation}
where $a,b$ encode the ratios of circular velocities. In the applet \url{https://www.geogebra.org/m/jugrcbx5}, their increment is defined to be 1.

\begin{figure}[ht]
\centering
\mbox{
\subfigure[5-fold]{\includegraphics[width=7.8cm]{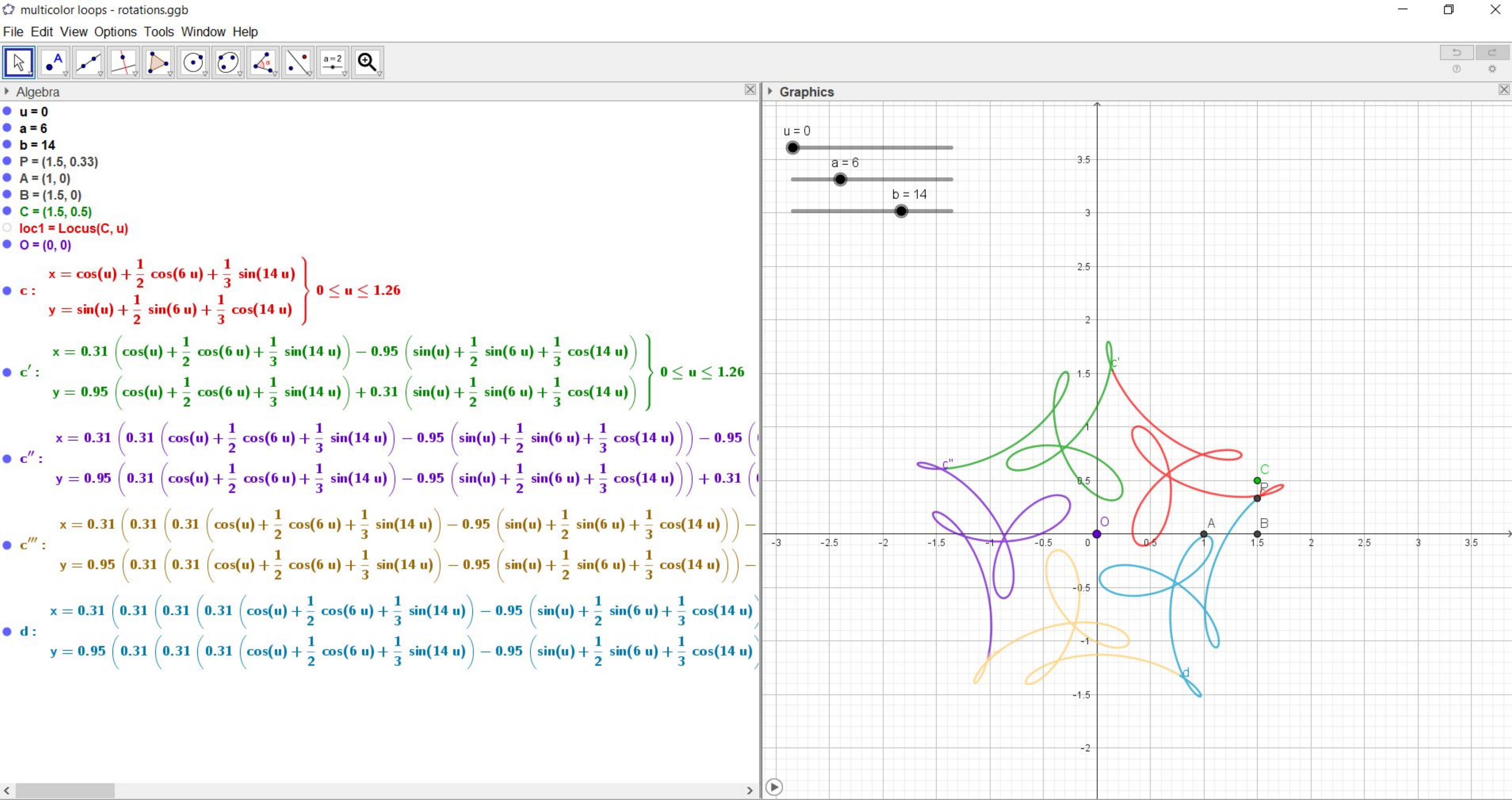}}
\quad
\subfigure[3-fold]{\includegraphics[width=5cm]{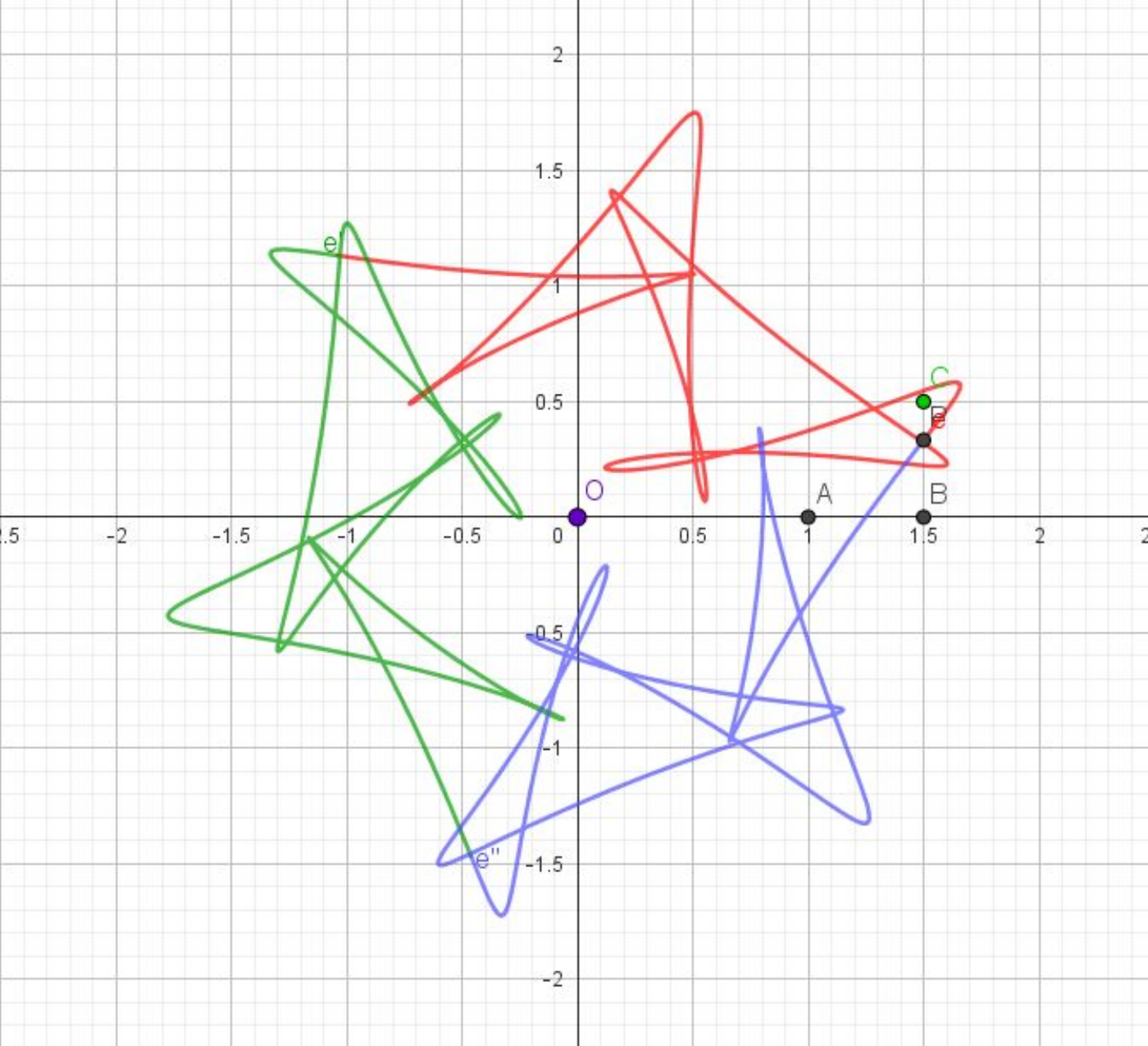}}
}
\caption{Tricircular moves creating multicolor curves with rotational symmetries}
\label{fig multicolor curves}
\end{figure}

For $a=b=1$, the curve is an ellipse. But there are other cases, maybe more interesting. Figure \ref{fig multicolor curves}(a) has been obtained for $(a,b)=(6,14)$ with the \textbf{Curve} command. It presents a 5-fold rotational symmetry, i.e. it is invariant under a rotation whose center is the origin and of angle $2 \pi /5$. This has been checked with a plot of the parametric equations for
 $u \in [0,2 \pi /5]$, the applying the automated command for rotations. The colors have to be manually adapted to create visualisations that are easier to interpret where each curve has a unique style. Part of GeoGebra's  algebraic display can be seen in  figure \ref{fig multicolor curves}(a) to illustrate what has been done. The  curve can also be plotted defining a variable point depending on the parameter $u$, then applying the \textbf{Locus} command. The definition of a variable point provides a dynamic plot of the curve, but both in this case and with the \textbf{Locus} command, the output is not a geometric object on which a plane transformation can be applied.

 Other cases have to be cautiously explored for symmetries. For example, the case $(a,b)=(10,14)$ shows a 3-fold symmetry (see figure \ref{fig multicolor curves}(b). Experomentation will show that this also true for $(a,b)=(7,14)$ and $(a,b)=(7,17)$.

\subsection{Math Art Creation}
\label{subsection math art}
\begin{figure}[ht]
\centering
\mbox{
\subfigure[]{\includegraphics[width=4.5cm]{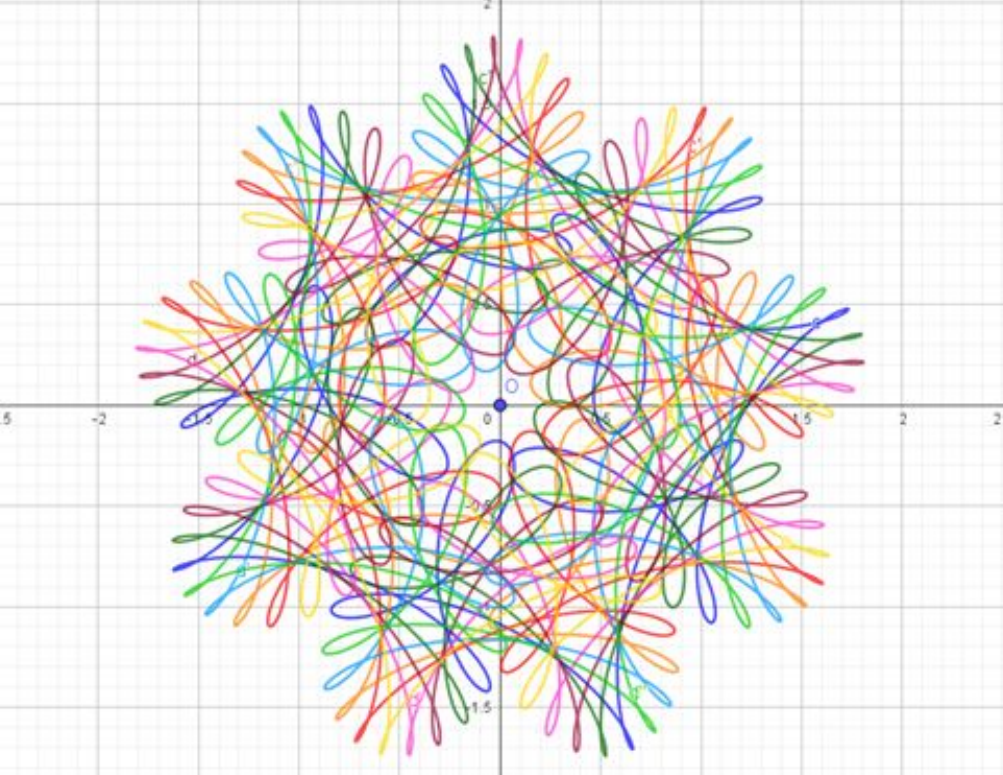}}
\quad
\subfigure[]{\includegraphics[width=4.5cm]{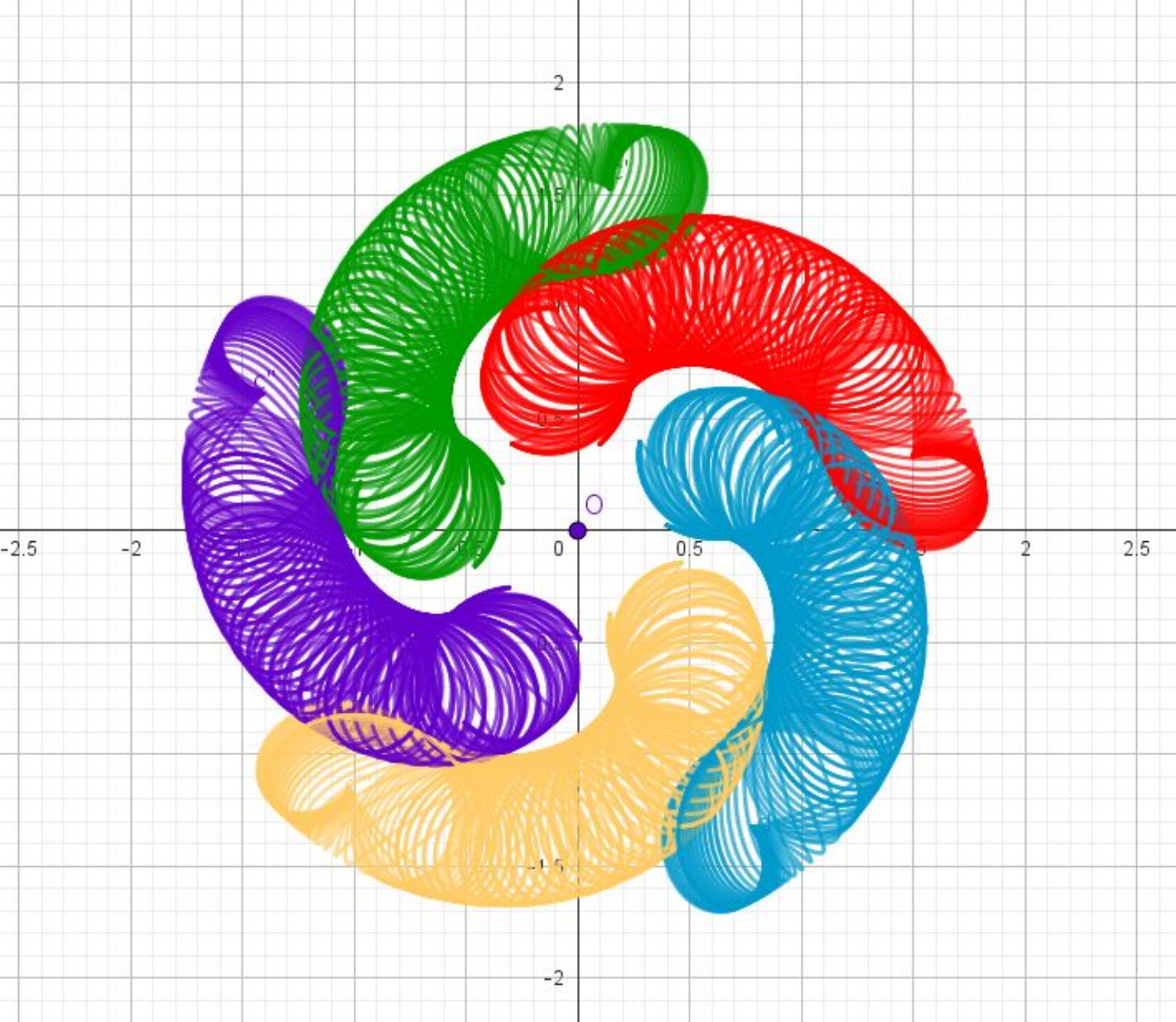}}
\quad
\subfigure[]{\includegraphics[width=4.5cm]{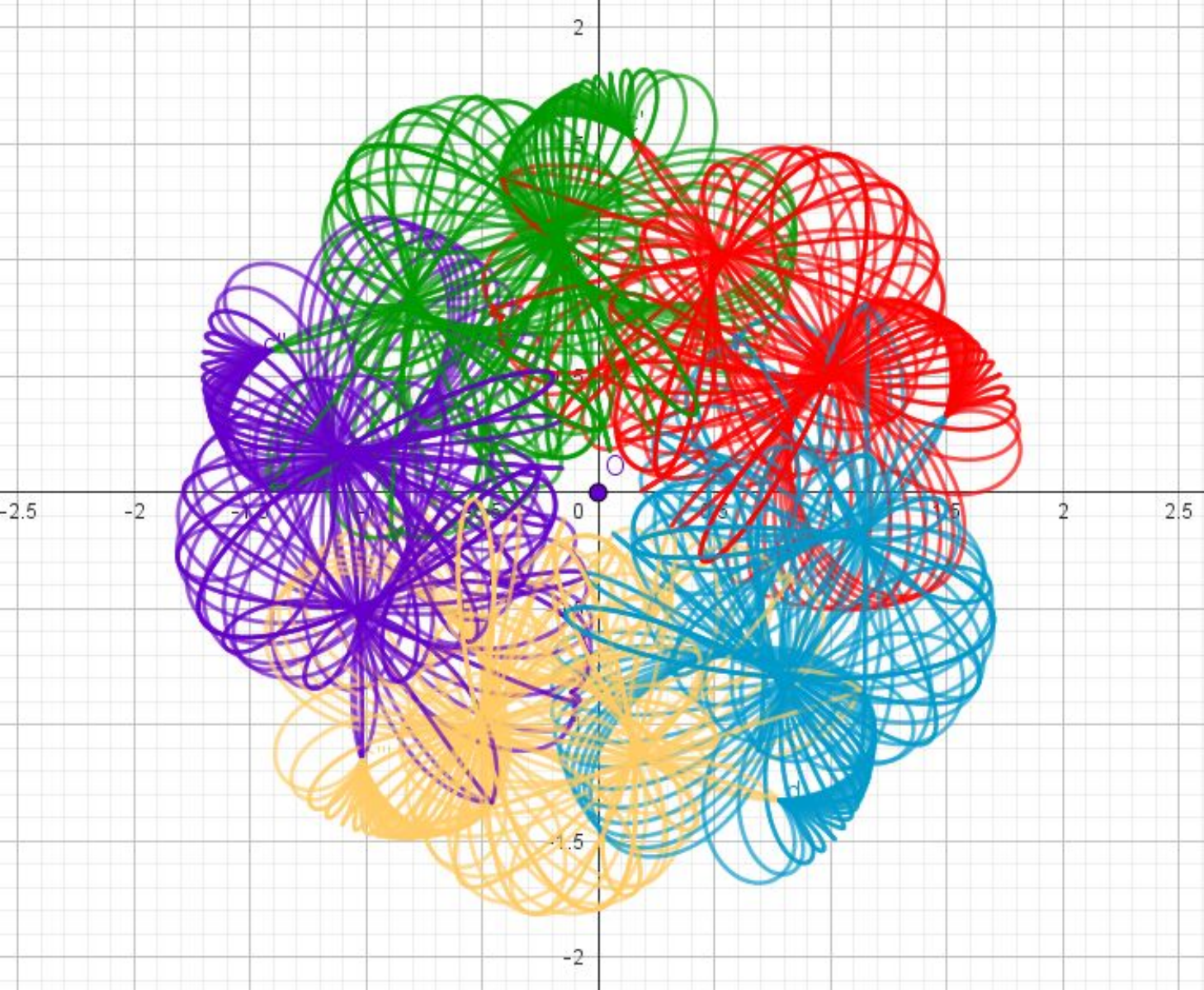}}
}
\caption{Some random math art creations}
\label{fig math art}
\end{figure}

The applet mentioned in the previous subsection has been opened, running animations for the parameters $a$ and $b$ separately. Exploration has been preformed according the following steps:
\begin{itemize}
\item The entire curve is plotted, using Trace On;
\item Analyzing the graphical display, the existence of rotational symmetry is conjectured;
\item The rotational symmetry is checked by first reducing the plot to a subset of the interval chosen for the parameter; we mean taking an interval of the form $[0, 2\pi / m]$, where $m$ is te order of the conjectured symmetry, and then using the automated command for a rotation about the origin with angle $2 \pi /m$.
\item Of course, this has to be checked afterwards by symbolic means, using a substitution.
\end{itemize}  
Later, an experiment has been made, choosing an arbitrary number $m$, not the order of the rotational symmetry which has been discovered. The obtained multicolor plot does not describe a specific mathematical situation. Some of the results are displayed in figure \ref{fig math art}.

Discovering such creations has been greeted with enthusiasm by the audience of lectures delivered by the authors, whose topics was linked to curves and math art.

\section{Some More Remarks}

The starting point of the study is STEAM oriented, namely using a scientific model from an item in the news. Students may have prior interest in the domain, without having a strong knowledge. The present topic offers an opportunity to collaborate between educators, between man and machine, of course between students. The study output is multiple, and among the ``rewards''   we have:
\begin{itemize}
\item Acquisition of new mathematical knowledge: classical curves (epicycloids, epitrochoids, etc.), which are not part of the regular curriculum, have been discovered and studied. Epitrochoids are members of a larger family of curves, which involves roses, epicycloids, etc. Activities as in this work may be a nice incitement to explore other situations and to broaden horizons. The literature describes generally the epitrochoids for integer values of the parameters, and our experimentations showed also more general settings.
\item Discovery of new curves; we mean curves which do not appear in the catalogues such as \cite{yates}.
\item Emphasis on the importance of the data precision (in space, contrary to most classrooms, nothing is measured by integers) and of rounding. We considered non integer ratios of radii of orbits, and of orbital angular velocity, approximations and rounding became an important issue. We could discover that different precisions in the approximation yield very different output. This is probably a central outcome of this work: students do not always believe that mastering errors is important, and they believe that the answers provided by a numerical calculator are always accurate.  Asking them which answer is true among the cases that we studied with different rounding should lead at least to some questioning.
\item Development on new technological skills, which are part of the new mathematical knowledge \cite{artigue}.
\item Emphasis on multidisciplinary tasks, whence development of  STEAM skills.
\end{itemize}

\begin{figure}[htb]
\begin{center}
\scalebox{0.25}{
\includegraphics{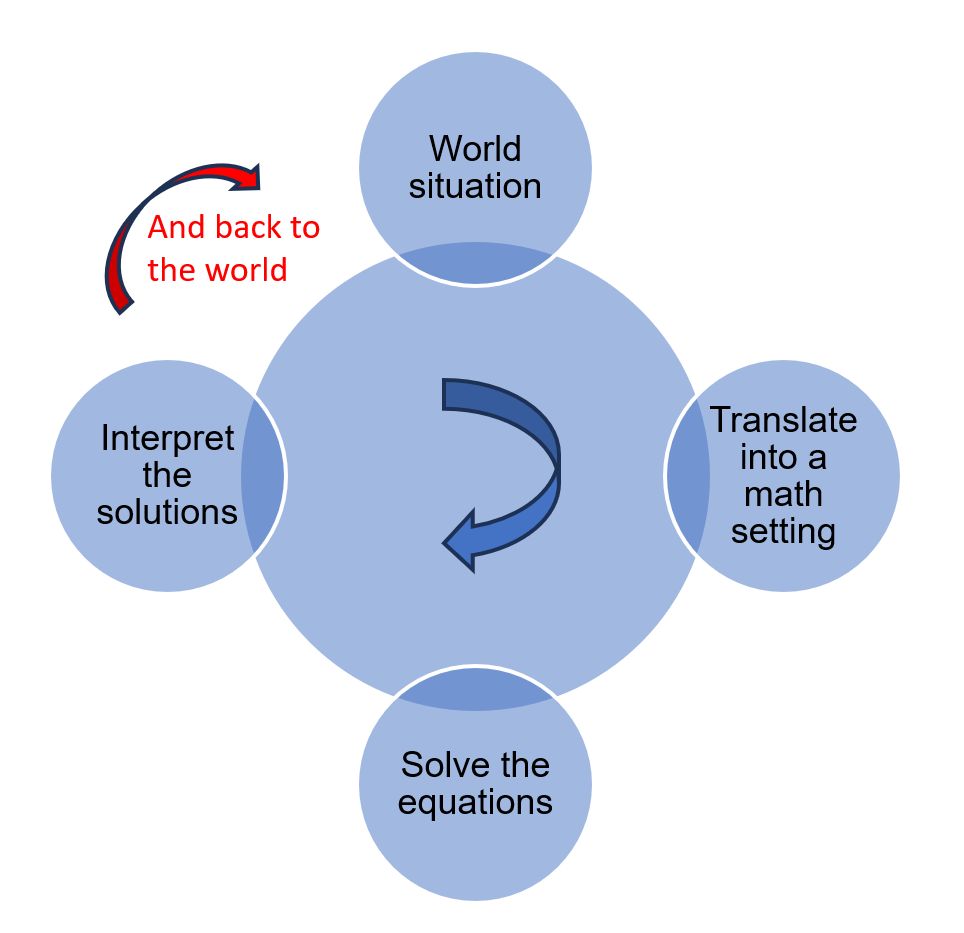}
}
\caption{A classical diagram for a modelling process}
\label{fig modelling diagram}
\end{center}
\end{figure}

Note that generally, modeling is intended to construct mathematical descriptions of a concrete situation. Then, the model is applied to enhance
more understanding of the concrete situation. The process is summarized in figure \ref{fig modelling diagram}.

\begin{figure}[htb]
\begin{center}
\scalebox{0.4}{
\includegraphics{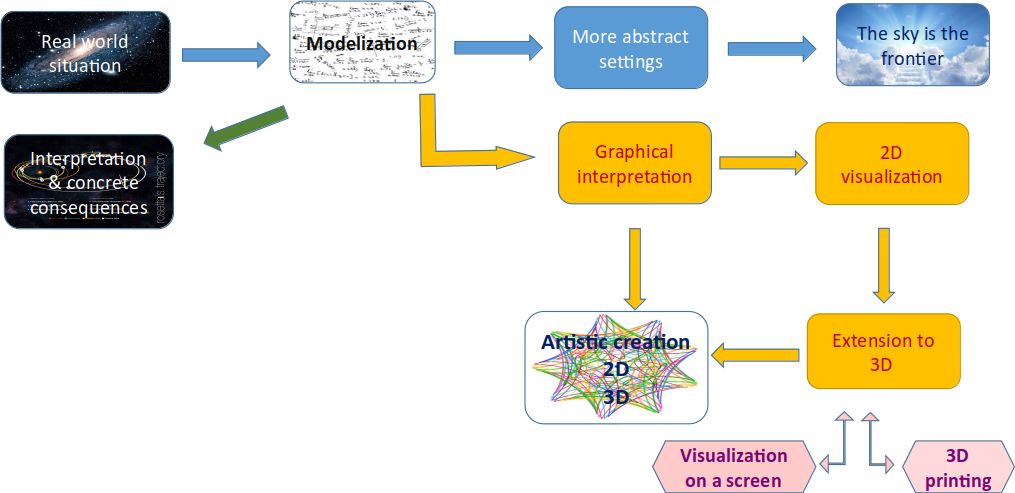}
}
\caption{A modelling process leading in other directions}
\label{fig extended modelling diagram}
\end{center}
\end{figure}

In the present paper, we go in a totally different way in this case: modeling a concrete astronomical situation (orbits), the activities
provide more abstract curves without a physical meaning. Finally 3D printing could provide both outcomes: a concrete object modeling planets and
trajectories, and also some pieces of visual art, either in 2D or in 3D to apply constructivist as well as constructionist ideas. This is summarized in figure \ref{fig extended modelling diagram}.

We performed the same experiments and constructions using Maple. The characteristics of the work  CAS is slightly different.
\begin{itemize}
\item After a command line to define a parametric curve, an \textbf{animate} command has to be entered. Its output is not immediately visible.
\item A left-click on the graphical window is necessary, and it switches automatically to the row of graphical buttons.
\item Here too, the relevant values for the parameters (number of frames, speed, etc.), in order to obtain a significant graphical output have to be experimentally looked for, using the buttons.
\item Other modifications of the output may require changes in the written commands.
\end{itemize}

After having presented some of the applets to a certain audience, the authors decided to 3D print part of them, together with some other cases. In parallel, tasks have been defined for groups of students, either gifted High-School students having benefit of an extension of the curriculum, or undergraduates. These students belong to two different countries. The tasks include the 3D printing of some examples. The transfer of the CAS output to a 3D printer requested the translation of this output into a language that the 3D printer understands.
In our presentation, we will report on the math part and on the outcome of the activities with students.

\section{Acknowledgements}
The first author was partially supported by the CEMJ Chair at JCT.

\bibliographystyle{eptcs}

\end{document}